\documentclass[%
 aip,
 amsmath,amssymb,
 reprint,%
]{revtex4-1}

\usepackage{graphicx}
\usepackage{dcolumn}
\usepackage{bm}
\usepackage[colorinlistoftodos]{todonotes}
\usepackage{subcaption}

\usepackage{xcolor}
\usepackage{amsmath}
\usepackage{xcolor}
\usepackage{bm}
\usepackage{amssymb}
\usepackage{wrapfig}
\usepackage{cancel}
\usepackage{soul} 
\usepackage{nicefrac}
\usepackage{hyperref}

\usepackage[utf8]{inputenc}
\usepackage[T1]{fontenc}
\usepackage{mathptmx}
\usepackage{etoolbox}
\newcommand{\unstretch}[1]{\mbox{#1}}
\newcommand{\edit}[1]{#1}
\makeatletter
\def\@email#1#2{%
 \endgroup
 \patchcmd{\titleblock@produce}
  {\frontmatter@RRAPformat}
  {\frontmatter@RRAPformat{\produce@RRAP{*#1\href{mailto:#2}{#2}}}\frontmatter@RRAPformat}
  {}{}
}%
\makeatother
\begin{document}

\preprint{AIP/123-QED}

\title[Computation of the Biot-Savart \edit{l}ine \edit{i}ntegral \edit{with higher-order convergence using straight segments}]{Computation of the Biot-Savart \edit{l}ine \edit{i}ntegral \\
\edit{with higher-order convergence using straight segments}}

\author{N. McGreivy}
\affiliation{Department of Astrophysical Sciences, Princeton University, Princeton NJ \edit{08540}, USA}
\affiliation{ Princeton Plasma Physics Laboratory, Princeton NJ 08543-0451, USA}
\email{mcgreivy@princeton.edu}
\author{C. Zhu}

\affiliation{ Princeton Plasma Physics Laboratory, Princeton NJ 08543-0451, USA}

\author{L. Gunderson}
\affiliation{Gatsby Computational Neuroscience Unit, University College London, \edit{WC1N 3AR}, United Kingdom}

\author{S. Hudson}
\affiliation{ Princeton Plasma Physics Laboratory, Princeton NJ 08543-0451, USA}


\date{\today}

\begin{abstract}
One common approach to computing the magnetic field produced by a filamentary current-carrying coil is to approximate the coil as a series of straight segments. The Biot-Savart field from each straight segment is known analytically. However, if the endpoints of the straight segments are chosen to lie on the coil, then the accuracy of the Biot-Savart computation is generally only second-order in the number of endpoints.
We propose a simple modification: shift each endpoint off the coil in the outwards normal direction by an amount proportional to the local curvature. With this modification, the Biot-Savart accuracy \edit{increases to fourth order} and the numerical error is dramatically reduced for a given number of discretization points.
\end{abstract}

\maketitle

\section{\label{sec:intro} Introduction}
\draft \edit{Students} of first-year college physics \edit{are taught} the Biot-Savart formula for the magnetic field produced by a \edit{static} current-carrying wire loop:
\begin{equation}\label{eq:filamentary_biot_savart}
    \bm B(\bm r) = \frac{\mu_0 I}{4\pi}\edit{\oint} \frac{ d\bm r' \times (\bm r - \bm r')}{|\bm r - \bm r'|^3}.
\end{equation}
These students learn to use the Biot-Savart law to compute the magnetic field produced by a variety of symmetric configurations such as straight segments and circular coils. \edit{Meanwhile, as researchers in plasma physics, we numerically compute the Biot-Savart law in a variety of applications including field line tracing,\cite{Pedersen2016} particle tracking, magnetohydrodynamic equilibrium codes,\cite{stellopt} and stellarator coil optimization codes.\cite{Zhu_2017}} But if the magnetic field cannot be calculated analytically, then equation \ref{eq:filamentary_biot_savart} must be computed numerically. Ideally, \edit{we would have a method for computing the Biot-Savart line integral that would be} simple enough to be taught to students of even first-year physics \edit{yet accurate and efficient enough to be used in research applications}. 

There are three main approaches to computing the Biot-Savart line integral given by equation \ref{eq:filamentary_biot_savart}. The first approach involves using a numerical quadrature and is usually highly accurate, but \edit{gives $\bm{\nabla}\times\bm{B}\ne 0$ and} requires storing the coil parameterization $\bm r'(s)$ or a series of position vectors $\bm r'_i$ and tangent vectors $\delta \bm r'_i$. The second approach, described in the next paragraph, is inaccurate \edit{relative to the first approach} but is conceptually simple and is widely used because it only requires storing a series of discrete points $\bm r'_i$ \edit{and because, to machine precision, $\bm\nabla \times \bm B =0$ when the evaluation point does not lie on the curve}. The third approach relies on the Fast Multipole Method\cite{greengard_rokhlin_1997} and is discussed in the related work section (section \ref{sec:relatedwork}). The purpose of this paper is to propose a new approach to evaluating the Biot-Savart line integral which is both highly accurate and conceptually simple \edit{while giving $\bm\nabla \times \bm B =0$},\footnote{\edit{It is well known that the Biot-Savart integral has zero curl in regions where $\bm J = 0$, as predicted by Ampere's law. However, the integrand of the Biot-Savart law has non-zero curl even where $\bm J = 0$. Therefore, in contrast with many other methods (e.g., quadrature, discontinuous piecewise linear), the explicit construction of a closed filamentary curve ensures that the resulting magnetic field is curl-free to machine precision.}} thereby enjoying the benefits of both the first and second approaches.

\begin{figure}
    \centering
	\begin{subfigure}[t]{0.5\textwidth}
		\centering%
		\includegraphics[width=\linewidth]{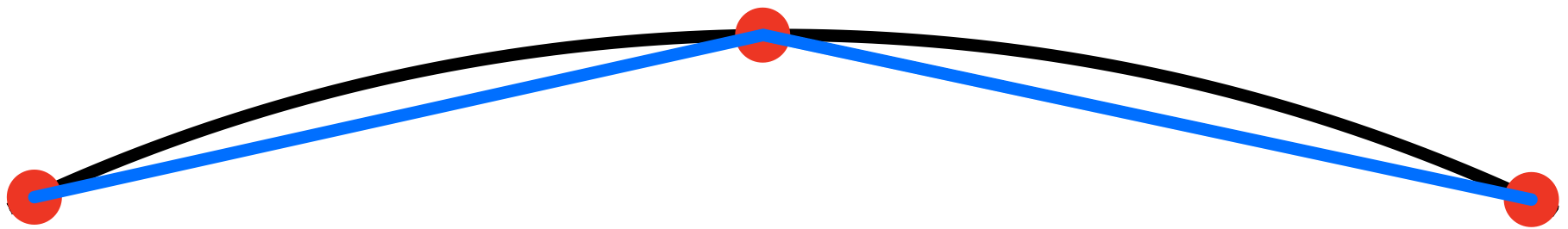}
    \caption{}
    \label{fig:piecewiselinear}
	\end{subfigure}
	\begin{subfigure}[t]{0.5\textwidth}	
		\centering%
		\includegraphics[width=\linewidth]{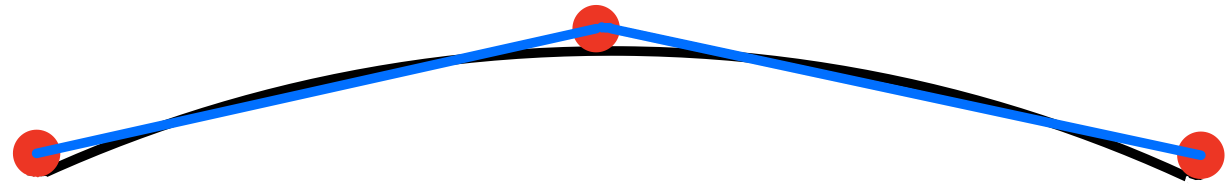}
		\caption{}
		\label{fig:shiftedpiecewiselinear}
	\end{subfigure}
	\caption{(a) Standard piecewise linear approach. Here the endpoints (red) lie on the filamentary curve (black). (b) Shifted piecewise linear approach. Here the endpoints are shifted in the outwards normal direction relative to the curve.}
	\label{fig:sketch}
\end{figure}

The second approach to numerically computing equation \ref{eq:filamentary_biot_savart} is to approximate the coil-carrying filament as a series of straight segments, then sum the magnetic field from each segment. As a convenient analytic expression is known for the magnetic field resulting from a straight segment\cite{Hanson_Hirshman}, a filamentary coil may be simply represented as an array of discrete endpoints. These endpoints lie on the coil and are typically equally spaced in distance or parameterization angle. This approach to computing equation \ref{eq:filamentary_biot_savart}, which we call the `standard piecewise linear approach', is illustrated in figure \ref{fig:piecewiselinear}. In figure \ref{fig:piecewiselinear}, the filamentary curve is shown in black, the discrete endpoints are shown in red, and the piecewise linear segments are shown in blue. 

Although the standard piecewise linear approach is convenient, simple, and flexible, it is not a particularly accurate way of computing the Biot-Savart line integral. One source of numerical inaccuracy can be understood from figure \ref{fig:piecewiselinear}: because the red endpoints lie on the curve, then the blue straight segments always lie locally inside the black curve. 
Because the straight segment always lies locally inside the curve, the field from the segment is biased and the result is that the accuracy of the Biot-Savart computation is only second-order in the number of endpoints.

We therefore propose a slightly different approach. Instead of choosing endpoints which lie on the filament, we choose endpoints which are slightly shifted in the outwards normal direction. This alternative approach, which we call the `shifted piecewise linear approach', is sketched in figure \ref{fig:shiftedpiecewiselinear}. Now the blue segments sometimes lie inside and sometimes lie outside the black curve, which eliminates the source of bias mentioned in the previous paragraph. The result is that, if the proper shift is chosen, then the accuracy of the Biot-Savart computation will be fourth order in the number of endpoints and the error will be dramatically reduced.

\begin{figure}
    \centering%
	\includegraphics[width=\linewidth]{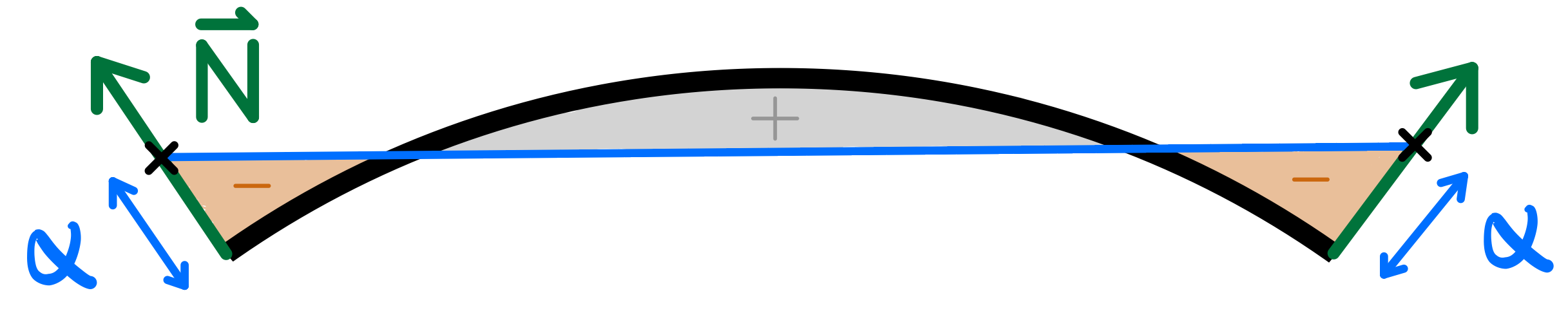}
	
	\caption{A close-up of the geometry of a shifted linear segment. The filamentary coil is shown in black. The normal vector to the curve $\bm N$ is drawn in green, and the straight segment is drawn in blue. The black crosses represent the endpoints of the straight segments, and the magnitude of the shift is given by $\alpha$. To lowest order, the area shaded in grey (positive) is equal to the area shaded in brown (negative) if $\alpha$ is given by equation \ref{eq:alpha}. }
	\label{fig:shift}
	
\end{figure}

It remains to determine by what magnitude to shift the endpoints in the outwards normal direction. Figure \ref{fig:shift} sketches a close-up of the geometry of a shifted linear segment, where the magnitude of the shift is labeled by $\alpha$. In appendix \ref{sec:appendixA}, we compare the Biot-Savart fields from a curved segment and shifted straight segment, and show that a shift of
\begin{equation}\label{eq:alpha}
    \alpha = \frac{\kappa |\delta\bm r'|^2}{12}
\end{equation}
in the outwards normal direction cancels the second-order error in the Biot-Savart computation near the center of the coil. Similarly, in appendix \ref{sec:appendixB} we show that the same shift minimizes the mean squared deviation between the curved segment and the straight segment. Here, the curvature \unstretch{$\kappa \equiv 1/R$}, where $R$ is the radius of curvature, \unstretch{$\delta \bm r' \equiv (d\bm r'/ds) \delta s$}, $s$ is a coordinate parametrizing the coil, $\delta s$ is the coordinate spacing between points, and the normal direction is defined by the Frenet-Serret formulas. 

\edit{Inspired by the calculations in appendices \ref{sec:appendixA} and \ref{sec:appendixB}, we propose choosing $\alpha$ using equation \ref{eq:alpha}. Although we have not discovered an analytic proof for the order of convergence with this shift, we do present careful numerical convergence studies (see section \ref{sec:tests}) to test our method. Based on these numerical convergence experiments, we conclude that once there are enough endpoints such that the curve between endpoints is locally well-described by a parabola, then the convergence will be fourth-order for a closed curve with continuous curvature. To ensure fourth-order convergence of a piecewise differentiable curve or non-closed curve requires a slightly modified approach, which is described in appendix \ref{sec:appendixD}. }

In summary, the shifted piecewise linear approach is a highly accurate method of computing the magnetic field produced by a twice-differentiable filamentary coil. The piecewise linear approach results in a \edit{continuous curve and a} piecewise constant representation of the tangent vector and therefore \edit{gives $\bm \nabla \times \bm B = 0$ and} only requires storing a series of discrete points $\bm r'_i$. This is in contrast to the numerical quadrature method, which \edit{gives $\bm \nabla \times \bm B \ne 0$ and which} requires storing a set of discrete points $\bm r'_i$ along with their associated tangent vectors $\delta \bm r'_i$.

\section{\label{sec:tests} Numerical Tests}

\draft In this section, we perform a few numerical tests of our proposed approach, the so-called `shifted piecewise linear approach'. To do so, we compare the Biot-Savart accuracy of the shifted piecewise linear approach to the accuracy of the standard piecewise linear approach as a function of the number of discretization points $N$ for three separate test problems: (1) an off-axis measurement from a circular coil, (2) an off-axis measurement from a non-planar modular stellarator-like coil, and (3) an on-axis measurement of the magnetic field from a D-shaped, tokamak-like coil. We measure the error as a function of the number of discretization points $N$, $\epsilon_N$, using the normalized $\ell^2$ norm between the produced magnetic field for a given number of discretization points, \unstretch{$\bm B_N(\bm r)$}, and the magnetic field as \unstretch{$N \rightarrow \infty$}, \unstretch{$\bm B_\infty (\bm r)$}:
\begin{equation}\label{eq:error}
    \epsilon_N = \frac{|\bm B_N (\bm r) - \bm B_\infty (\bm r)|}{|\bm B_\infty (\bm r)|}.
\end{equation}
To avoid possible biases related to the particular choice of measurement point $\bm r$, in each test problem we average the error $\epsilon_N$ over a set of $\edit{R=100}$ measurement points $\{\bm r_j\}_{j=1}^R$ where \unstretch{$\bm r_j = \bm r + \bm \delta_j$} and $\bm \delta_j$ is a small random vector. \edit{In} each test \edit{problem}, the \edit{maximum deviation of $\epsilon_N$ relative to the average is less than an order of magnitude}. \edit{For test problem 1, $\bm B_\infty$ is known analytically and involves a computationally efficient elliptic integral.\cite{urankar_part1} }

\begin{figure*}
\centering
\includegraphics[width=\linewidth]{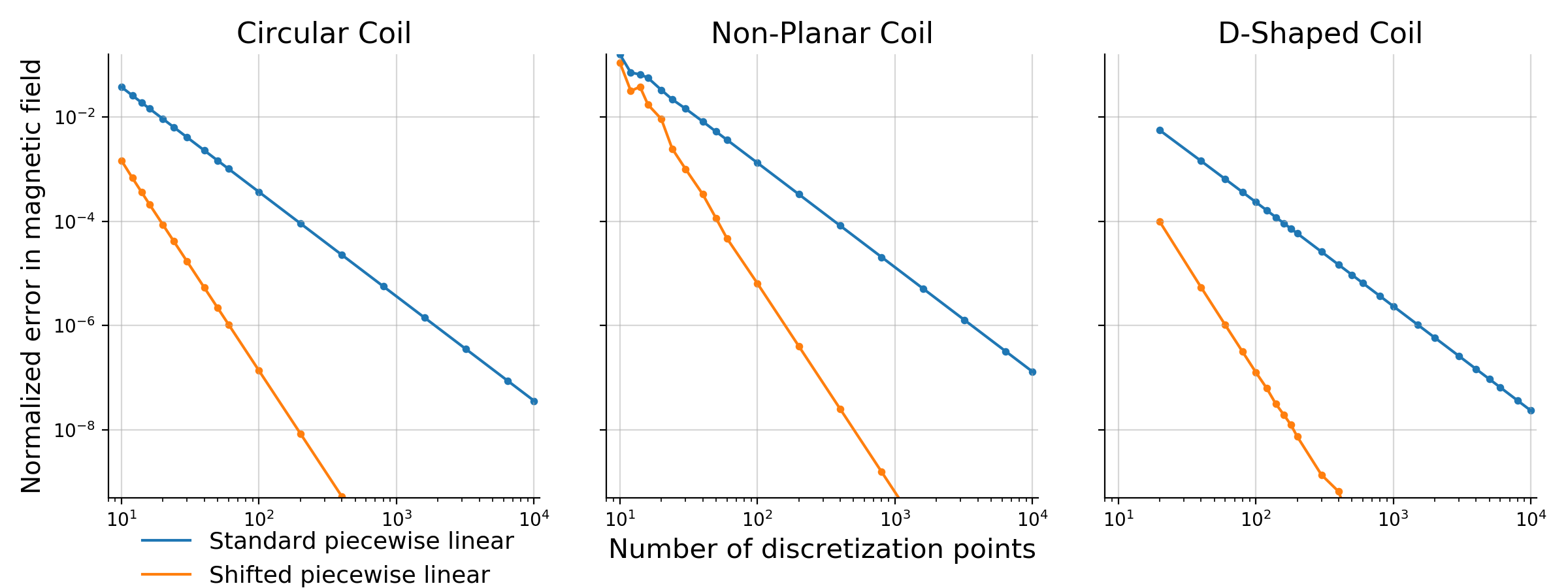}
\caption{The average error $\epsilon_N$ as a function of the number of discretization points $N$ for (left) a circular coil (center) a stellarator-like, non-planar coil, and (right) a tokamak-like, D-shaped coil. The shifted piecewise linear approach is fourth-order in each experiment, while the standard piecewise linear approach is second-order. \edit{Code for these experiments, as well as additional experiments, can be found at} \href{https://github.com/nickmcgreivy/biot-savart-line-integral}{\textcolor{blue}{https://github.com/nickmcgreivy/biot-savart-line-integral}}.}
\label{fig:results}
\end{figure*}

Our results are shown in figure \ref{fig:results}. In each case, $\epsilon_N$ is smaller using the shifted piecewise linear approach by at least an order of magnitude compared to the standard piecewise linear approach for all \unstretch{$N > 20$}. Additionally, while the accuracy of the standard piecewise linear approach is second-order, the accuracy of the shifted piecewise linear approach is fourth-order for each experiment. The shifted piecewise linear approach is significantly more accurate than the standard piecewise linear approach for a given number of discretization points.

We also \edit{have found} that our results are \edit{qualitatively} robust to the choice of measurement point $\bm r$: it doesn't matter whether $\bm r$ is on-axis, off-axis, near the coil or far from the coil, inside or outside the coil. In each case, the \edit{convergence of the shifted piecewise linear approach remains fourth-order}.

\section{\label{sec:relatedwork} Related Work}
\draft Early work by \citeauthor{urankar_part1} 
has developed analytic expressions for the magnetic field produced by filamentary circular\cite{urankar_part1} and elliptic\cite{urankar_elliptic} arc segments, thin conic current sheets \cite{urankar_part2}, rectangular cross-section circular arcs with azimuthal\cite{urankar_part3} and arbitrary\cite{urankar_part4} current densities, and polygonal cross-section circular arcs\cite{urankar_part5}. \citeauthor{Hanson_Hirshman} present new compact expressions for the analytic fields from a straight line segment.\cite{Hanson_Hirshman} \citeauthor{suh_2000_surfaceintegral} develops closed-form expressions for the magnetic field from volumetric elements with planar edges and constant and/or linear current density which involve line integrals over the boundary edges of the element.\cite{suh_2000_surfaceintegral} \citeauthor{integration_methods} examine a variety of methods for computing the Biot-Savart fields from a volumetric  current-carrying coil, including (i) splitting the domain into tetrahedral regions and performing a quadrature on each domain, and (ii) numerically integrating a semi-analytic one-dimensional integral based on Urankar's analytic expressions.\cite{integration_methods} \citeauthor{rectangular_prism_and_curved_segments} generate analytic expressions for the magnetic field from trapezoidal prisms.\cite{rectangular_prism_and_curved_segments} \citeauthor{nunes_curvilinear_elements} approximate filamentary, surface current, and volumetric current densities as first-order or second-order finite elements, then use a quadrature to evaluate the current from each finite element and an adaptive method which reduces the required number of Gauss points.\cite{nunes_curvilinear_elements}

\citeauthor{magnetic_field_of_modular_coils} rewrite the Biot-Savart volume integral for a rectangular cross-section finite-build modular stellarator coil as a simpler, tractable integral over finite-build coil coordinates.\cite{magnetic_field_of_modular_coils} Although the integrals over the coil cross-section have no analytic solution, they can either be evaluated numerically or semi-numerically in a power series. \citeauthor{mcgreivy2020optimized} have approximated the same integral using a multi-filament approximation and a quadrature-based approach to computing the Biot-Savart line integral.\cite{mcgreivy2020optimized} The approach of using a numerical quadrature requires computing the tangent vector to the filamentary coil; this approach is often also significantly more accurate than the standard piecewise linear approach but is not examined in this paper to simplify the presentation.

A different approach to computing the Biot-Savart law uses the Fast Multipole Method (FMM) to improve the runtime scaling of the Biot-Savart law from \unstretch{$\mathcal{O}(NM)$} to \unstretch{$\mathcal{O}(N+M)$} where $N$ is the number of sources and $M$ is the number of evaluation points. This approach to computing the Biot-Savart law will significantly improve the runtime of a Biot-Savart \edit{calculation} which requires a large number of evaluation points $M$, but a runtime comparison of these approaches as $N$ and $M$ vary is outside of the scope of this paper. Recently, \citeauthor{ProjectRat} have developed an open-source library called Project Rat which implements this approach.\cite{ProjectRat}

\section{\label{sec:conclusion} Conclusion}

\draft We investigated the standard piecewise linear approach to computing the magnetic field from a filamentary coil. We found that this approach is only second-order accurate if the endpoints lie on the coil, because the straight segments are biased to always lie locally inside the coil. We propose a new approach, the shifted piecewise linear approach, where the endpoints are shifted in the outwards normal direction by an amount proportional to the local curvature. We find that a shift of \unstretch{$\kappa |\delta\bm r'|^2/12$} minimizes the squared distance in real space between the filament and the straight segment, while also cancelling the second-order errors in the Biot-Savart field near the radius of curvature of the filament. We conduct simple numerical tests, and find that the shifted piecewise linear approach has higher-order convergence than the standard piecewise linear approach and is dramatically more accurate for a given number of discretization points.

There are two main conclusions. First, the shifted piecewise linear approach is superior to the standard piecewise linear approach, and should be used when possible. Second, when filamentary coils are stored in databases, the stored endpoints should not lie on the coil but rather should be shifted normally outwards. 
\begin{acknowledgments}
This work was supported by the U.S. Department of Energy under Contract No. DE-AC02-09CH11466 through the Princeton Plasma Physics Laboratory.
\end{acknowledgments}

\section*{Data Availability Statement}

The data that support the findings of this study are available from the corresponding author upon reasonable request.

\appendix
\section{Derivation of Piecewise Linear Shift \label{sec:appendixA}}
Here we show that by shifting each point on the filamentary coil in the outwards normal direction by an amount \unstretch{$\kappa|\delta \bm r'|^2/12$}, where $\kappa$ is the local curvature and \unstretch{$d \bm r' \equiv (d\bm r'/ds) \delta s$}, cancels the second-order error in the Biot-Savart field near the center of the coil. To do this, we compare the Biot-Savart fields for the straight segment and curved section as shown in figure \ref{fig:shift}. \edit{With the origin defined as the midpoint between the two endpoints of the curved segment in figure \ref{fig:shift},} the straight segment can be written as \unstretch{$\bm r' = x' \hat{x} + \alpha \hat{y}$} for \unstretch{$x'\in [-L/2(1 + \alpha/R), L/2(1 + \alpha/R)]$} while the curved segment can be written as \unstretch{$\bm r' = x' \hat{x} + ({L^2}/{8R} - {x'^2}/{2R}) \hat{y}$ for $x'\in [-L/2, L/2]$}. \edit{This quadratic parametrization of the curved segment is equal, through second order, to that of any arbitrary smooth curve.} \edit{We compute the magnetic field at the point \unstretch{$\bm r_0 = (0, -R, 0)$} where \unstretch{$R \equiv 1/\kappa$}.} We compute the Biot-Savart fields for the straight segment and curved segment and work in the limits \unstretch{$L/R \sim \alpha/L \sim \mathcal{O}(\epsilon)$} and \unstretch{$\alpha/R \sim \mathcal{O}(\epsilon^2)$}, keeping all terms of up to second order.

The magnetic field for the straight segment at $\bm r_0$ is given by the Biot-Savart law (equation \ref{eq:filamentary_biot_savart}) and equal to
\begin{align}
    \bm B(\bm r_0) =&\nonumber\\ -&(R + \alpha) \hat{z} \int_{-L/2(1 + \alpha/R)}^{L/2(1 + \alpha/R)} \frac{dx'}{\Big(x'^2 + (R + \alpha)^2\Big)^{3/2}}.
\end{align}
Defining \unstretch{$\beta \equiv x'/ L$}, this can be written as
\begin{align}
    \bm B(\bm r_0) =&\nonumber \\ -\frac{L}{R^2}&(1+\frac{\alpha}{R}) \hat{z} \int_{-(1+\frac{\alpha}{R})/2}^{(1+\frac{\alpha}{R})/2} \frac{d\beta}{\Big((1 + \frac{\alpha}{R})^2 + \frac{L^2}{R^2}\beta^2\Big)^{3/2}}.
\end{align}
The integral can be performed analytically, it gives
\begin{equation}
        \bm B(\bm r_0) = -\frac{L}{R^2} \frac{\hat{z}}{\sqrt{\frac{L^2}{4R^2}(1 + \frac{\alpha}{R})^2 + (1 + \frac{\alpha}{R})^2}}.
\end{equation}
Expanding the denominator gives
\begin{equation}\label{eq:straightfinal}
    \bm B(\bm r_0) = -\hat{z}\frac{L}{R^2}\Big(1 - \frac{\alpha}{R} - \frac{L^2}{8R^2}\Big) + \mathcal{O}(\epsilon^4).
\end{equation}

The magnetic field for the curved segment is given by
\begin{equation}
    \bm B(\bm r_0) = -\hat{z}\int_{-L/2}^{L/2} \frac{R +  \frac{L^2}{8R} + \frac{x'^2}{2R}}{\Big(x'^2 + (R + \frac{L^2}{8R} - \frac{x'^2}{2R})^2\Big)^{3/2}}dx'.
\end{equation}
Again defining \unstretch{$\beta \equiv x'/ L$}, this can be written as
\begin{align}
    \bm B(\bm r_0) =&\nonumber \\ -\hat{z}\frac{L}{R^2}& \int_{-1/2}^{1/2} \frac{1 + \frac{L^2}{8R^2}(1 + 4\beta^2)}{\Big(\frac{L^2}{R^2}\beta^2 + (1 + \frac{L^2}{8R^2}(1 - 4\beta^2))^2 \Big)^{3/2}}d\beta.
\end{align}
Expanding the denominator, this gives
\begin{align}
    \bm B(\bm r_0) = &-\hat{z}\frac{L}{R^2} \int_{-1/2}^{1/2}\Big(1 + \frac{L^2}{8R^2}(1 + 4\beta^2) \nonumber\\&- \frac{3L^2}{2R^2}\beta^2 - \frac{3L^2}{8R^2}(1 - 4\beta^2)\Big) d\beta
\end{align}
\begin{equation}\label{eq:curvedfinal}
    \bm B(\bm r_0) = -\hat{z}\frac{L}{R^2}\Big(1 - \frac{5L^2}{24R^2} \Big) + \mathcal{O}(\epsilon^4)
\end{equation}

Comparing equations \ref{eq:straightfinal} and \ref{eq:curvedfinal}, we see that setting \unstretch{$\alpha = L^2/12R$} gives the same second-order error in $B_z$ at \unstretch{$\bm r_0 = (0, -R, 0)$}. Here, \unstretch{$R \equiv 1/\kappa$} and to fourth order \unstretch{$L^2 = |\delta \bm r'|^2 \equiv |(d \bm r'/ds) \delta s|^2$}.

\section{Minimization of Mean Squared Distance \label{sec:appendixB}}

With the same parameterization as above, we have that the squared distance from the curved segment to the straight segment is
\unstretch{$( {\alpha x'}/R )^2 + ( {L^2}/{8R} - {x'^2}/{2R} - \alpha )^2$} for \unstretch{$x'\in [-L/2, L/2]$}.  Taking the average over the segment gives
\begin{align}
    \langle |\Delta \bm r'|^2 \rangle = \int_{-1/2}^{1/2} \Big( \big( \frac{L^2}{8R}& - \alpha \big)^2 + \nonumber \\ \big( \frac{\alpha^2}{R^2} - \frac{1}{R}\big( &\frac{L^2}{8R} - \alpha \big) \big) L^2 \beta^2 + \frac{1}{4R^2} L^4 \beta^4 \Big) d\beta \nonumber\\
    = \Big( \frac{L^4}{64R^2} - \frac{L^2}{4R} \alpha + \alpha^2 \Big)&\nonumber \\ + \frac{1}{12}\Big( - \frac{L^4}{8R^2} + &\frac{L^2}{R} \alpha + \frac{L^2}{R^2} \alpha^2 \Big) + \frac{1}{80} \Big( \frac{L^4}{4R^2} \Big). \label{Eq:AverageSquaredDistance}
\end{align}
The derivative with respect to $\alpha$, 
\begin{equation}
    d\langle |\Delta \bm r'|^2 \rangle/d\alpha = - \frac{L^2}{4R} + 2\alpha + \frac{L^2}{12R} + \frac{L^2}{6R^2}\alpha, 
\end{equation}
when set to zero, gives the optimal shift of the straight segment: 
\begin{equation}
    \alpha = \frac{1}{12}\frac{L^2}{R} \Big( 1 + \mathcal{O}(\epsilon^2) \Big) \label{Eq:SquaredShiftDerivation}
\end{equation}


\section{Prescription for \edit{Piecewise Continuous} Curvature \label{sec:appendixD}}
If the coil curvature is only \textit{piecewise} continuous, one should approximate each \edit{segment} individually. \edit{Although we have found that closed curves with continuous curvature have fourth-order error convergence using the shifted piecewise linear approach, to obtain fourth-order convergence from a non-closed segment requires a slightly modified method. We describe this method below.}

Consider approximating the magnetic field coming from one such section, parameterized by $s_i\in[0,1]$ for $i=0,1,\ldots,N$.  The procedure is the same as before, with the following modulations: 
\begin{enumerate}
    \item Set $s_0=0$ and $s_N=1$.  These points will not be shifted.  
    \item For $i\in1,2,\ldots,N-1$, we used to place $s_i$ uniformly.  Instead, spread them out slightly, such that the spacing in $s$ between ($s_0$ and $s_1$) and ($s_{N-1}$ and $s_N$) are smaller than the others by a factor of $\sqrt{2}$.  
    \item Shift the points $1,2,\ldots,N-1$ using equation \eqref{eq:alpha}.  For all shifts, use the $\delta s$ between the interior points, not the shorter length of the endpoints.  
\end{enumerate}
%

\end{document}